\documentclass{ws-ijgmmp}

\usepackage{slashed}

\newcommand{\be}{\begin{equation}}\newcommand{\ee}{\end{equation}}
\newcommand{\bea}{\begin{eqnarray}}\newcommand{\eea}{\end{eqnarray}}
\newcommand{\brr}{\begin{array}}\newcommand{\err}{\end{array}}
\newcommand{\bit}{\begin{itemize}}\newcommand{\eit}{\end{itemize}}
\newcommand{\ben}{\begin{enumerate}}\newcommand{\een}{\end{enumerate}}

\newcommand{\ba}{\begin{array}}
\newcommand{\ea}{\end{array}}

\def\lf{\left}

\def\ri{\right}
\def\al{\alpha}

\def\la{\lambda}

\def\1{{_{1}}}\def\2{{_{2}}}

\def\noHe0{:\;\!\!\;\!\!:H_e(0):\;\!\!\;\!\!:}
\def\noHm0{:\;\!\!\;\!\!:H_\mu(0):\;\!\!\;\!\!:}

\def\lf{\left}

\def\ri{\right}

\def\al{\alpha}

\def\la{\lambda}

\def\1{{_{1}}}\def\2{{_{2}}}

\begin{document}

\markboth{Blasone, Capozziello, Lambiase, Petruzziello}
{Violation of Equivalence Principle in Extended Gravity}

%
\catchline{}{}{}{}{}
%

\title{VIOLATION OF THE EQUIVALENCE PRINCIPLE IN CURVATURE BASED EXTENDED GRAVITY AT FINITE TEMPERATURE}

\author{MASSIMO BLASONE}
\address{Dipartimento di Fisica, Universit\`a degli Studi di Salerno,
Via Giovanni Paolo II, 132 I-84084 Fisciano (SA), Italy\\
Istituto Nazionale di Fisica Nucleare, Sezione di Napoli, Gruppo collegato di Salerno, 
Via Giovanni Paolo II, 132 I-84084 Fisciano (SA), Italy\\
\email{blasone@sa.infn.it} }

\author{SALVATORE CAPOZZIELLO}
\address{Universit\`a degli studi di Napoli Federico II, Dipartimento di Fisica Ettore Pancini, Complesso Universitario di Monte S. Angelo, Via Cintia Edificio 6, 80126 Naples, Italy\\
Scuola Superiore Meridionale, Largo S. Marcellino 10, 80138 Naples, Italy\\
Istituto Nazionale di Fisica Nucleare, Sezione di Napoli, Complesso Universitario di Monte S. Angelo, Via Cintia Edificio 6, 80126 Naples, Italy\\
Tomsk State Pedagogical University, ul. Kievskaya, 60, 634061 Tomsk, Russia
\email{capozziello@unina.it} }

\author{GAETANO LAMBIASE}
\address{Dipartimento di Fisica, Universit\`a degli Studi di Salerno,
Via Giovanni Paolo II, 132 I-84084 Fisciano (SA), Italy\\
Istituto Nazionale di Fisica Nucleare, Sezione di Napoli, Gruppo collegato di Salerno, 
Via Giovanni Paolo II, 132 I-84084 Fisciano (SA), Italy\\
\email{lambiase@sa.infn.it} }

\author{LUCIANO PETRUZZIELLO}
\address{Dipartimento di Ingegneria Industriale, Universit\`a degli Studi di Salerno,
Via Giovanni Paolo II, 132 I-84084 Fisciano (SA), Italy\\
Istituto Nazionale di Fisica Nucleare, Sezione di Napoli, Gruppo collegato di Salerno, 
Via Giovanni Paolo II, 132 I-84084 Fisciano (SA), Italy\\
\email{lupetruzziello@unisa.it} }
\maketitle

\begin{history}
\received{(Day Month Year)}
\revised{(Day Month Year)}
\end{history}

\begin{abstract}
We review the possible violation of the Equivalence Principle at finite temperature $T$ in the framework of curvature based Extended Theories of Gravity. Specifically, we first show how it is possible to derive Equivalence Principle  violation from Quantum Field Theory at $T\neq0$. Subsequently, we exhibit how this  result can be precisely recovered by following an alternative path that envisages the employment of  generalized Einstein equations with a temperature-dependent energy-momentum tensor. Finally, we adopt the latter formalism in the context of some Extended  Gravity models to quantify the amount of Equivalence Principle violation. Specifically, Brans-Dicke Theory, Standard Model Extension and Conformal Gravity are considered in details.
\end{abstract}

\keywords{Equivalence Principle; Extended  Gravity;  Quantum Field Theory.}

\section{Introduction}

According to Einstein's original formulation~\cite{mtw}, it is always possible to locally simulate the effects of gravity by suitably tuning the magnitude of the acceleration. In particular, if with $\phi_g$ we denote the gravitational potential, we observe that
\be\label{aeg}
\textbf{a}=-{\boldsymbol{\nabla}}\phi_g
\ee
is the acceleration a system  experiences to mimic the presence of gravity in a sufficiently small region of space. Note that, in writing down Eq.~\eqref{aeg}, we have tacitly assumed a perfect correspondence between the inertial mass $m_i$ and the gravitational mass $m_g$. Such a statement is another way of addressing the Equivalence Principle (EP), which is considered by Einstein himself as the ``happiest thought'' of his life.  
On the other hand, it must be said that, since the aforementioned seminal idea, many other EP formulations have been proposed. To have a further insight on this topic, we succinctly give a panorama of all the currently affordable versions\footnote{In the present Section and henceforth, when we talk of EP we refer to its classical formulation. For recent investigations centered around its behavior in the quantum domain, see Refs.~\cite{zy}.}; for this purpose, we closely follow the approach of Refs.~\cite{will,willb,libe}.

\begin{description}
\item[Newton (Galilei) Equivalence Principle (NEP):] This principle has already been commented after Eq.~\eqref{aeg}. As a matter of fact, this axiom affirms that, in the Newtonian limit, $m_i\equiv m_g$. An equivalent way to rephrase it is related to the universality of gravity, which acts indiscriminately on any form of matter and energy.

\item[Weak Equivalence Principle (WEP):] \emph{In a gravitational field, the motion of test particles with negligible self-gravity does not depend on their physical features} 

In order to quantify self-gravity, we can define the dimensionless parameter
\be\label{dimp}
\lambda=\frac{Gm}{rc^2}\,,
\ee
with $G$ being the Newton constant, $c$ the speed of light, $m$ the mass of the test body and $r$ its linear size. Eq.~\eqref{dimp} is the ratio between the gravitational and the rest energy provided that NEP holds. Therefore, we can assert that, as long as $\la\ll 1$, self-gravity can be safely neglected.
In other words, this notion can be easily visualized by imagining that, if two test particles have the same initial conditions, they travel along the same geodesic, regardless of their properties (i.e. mass, charge, etc.). 

\item[Gravitational Weak Equivalence Principle (GWEP):] \emph{In a gravitational field and in vacuum, the motion of test particles does not depend on their physical features} 

Differently from the previous statement, we have relaxed the condition on self-gravity, which means that GWEP $\to$ WEP as $\la\to 0$. However, a similar modification entails a further requirement, which is identified with the presence of vacuum. This aspect turns out to be crucial, as the gravitational field of test bodies interacts with the physical environment in which they are embedded. In so doing, by virtue of the action-reaction principle, they would undergo a net force that would undermine the universality of the gravitational interaction, as this interaction would be a function of the test particles' properties. 

\item[Einstein Equivalence Principle (EEP):] \emph{The presence of a gravitational field does not affect fundamental non-gravitational physical tests locally and in any point of spacetime} 

By ``fundamental physical tests'' we refer to experiments that deal with the equations describing the behavior of single particles, thus excluding the ones which can be derived from them. For instance, the rules that govern the motion of  composite systems are not contemplated by EEP, as for such systems local gravitational effects may be detected via experimental tests. On the other hand, by ``in any point of spacetime'' we mean that there are no privileged points on the spacetime manifold. Consequently, it is not important to learn where and when in the universe the experiment is performed, since the outcome shall not depend on it. This concept underlies the so-called local position invariance~\cite{will}. 
Therefore, we notice that EEP is the simultaneous requirement of WEP, local Lorentz invariance and local position invariance~\cite{will,willb}.

\item[Strong Equivalence Principle (SEP):] \emph{The presence of a gravitational field does not affect all fundamental physical tests (including gravitational physics) locally and in any point of spacetime}

Insofar, from its definition, it should be clear that SEP is the simultaneous requirement of gravitational weak EP, local Lorentz invariance and local position invariance. Hence, according to SEP, it is possible to perform even gravitational local experiments in presence of an external gravitational field, with the results not being invalidated by that.
\end{description}
For more details on this topics, we remand the interested reader to Refs.~\cite{will,willb,libe,weprev}.

In this work, we are mainly concerned with the WEP formulation, which is thoroughly studied in the framework of both General Relativity (GR) and Extended Theories of Gravity: here we intend extensions of GR where the Einstein theory is a particular case of wide classes of curvature based theories (for reviews, see Refs.~\cite{cap}, for applications Refs.~\cite{original,app}).  In the former scenario, the role played by the non-vanishing temperature $T$ is fundamental to detect a violation of the WEP. Indeed, we will demonstrate how, at zero temperature, EP still holds for GR, since the thermal corrections to the inertial and the gravitational mass vanish when $T\rightarrow 0$. More details on the thermal nature of the violation of the equivalence principle in close connection with the explicit breaking of the local Lorentz symmetry can be found in Ref.~\cite{gasper}. On the contrary, when working with Extended Theories of Gravity, WEP can be violated also when $T=0$. This means that further geometrical degrees of freedom can play a main role in the dynamics of EP violation. 

In light of the above discussion, the manuscript is organized as follows: in Sec.~II we review the investigation of EP violation in the context of Quantum Field Theory (QFT) at finite temperature. Section~III is devoted to the same analysis that however has a different starting point, since the attention is focused on the modification of the Einstein field equations. By exploiting the latter approach, we show that it is a simple task to extend the lines of reasoning for the Schwarzschild solution to other physical settings, whose spacetime can be described by means of Extended Theories of Gravity, which instead will be treated in Sec.~IV. Section~V contains conclusion and discussions.

Throughout the whole paper, we use the units $c=\hbar=k_B=1$ and the mostly-negative metric signature $\mathrm{diag}(+,-,-,-)$.

\section{Equivalence Principle violation: a first approach}

In order to deal with EP violation in the context of QFT, we closely follow the method explored for the first time in Ref.~\cite{don} (for modified gravity, see Refs.~\cite{hui}, whilst for the generalized Uncertainty Principle see Ref.~\cite{sc}).
The system under consideration is made up of an electron with renormalized mass $m_{0}$ at zero temperature which is in thermal equilibrium with a photon heat bath. The main goal of the whole investigation lies in the evaluation of the electron's gravitational and inertial mass
in the low-temperature limit $T\ll m_{0}$. The presence
of a non-vanishing temperature is crucial, as it will be explicitly
shown that $m_{g}=m_{i}$ for $T=0$.
The gravitational and inertial mass are deduced by resorting to a Foldy\textendash Wouthuysen
transformation~\cite{fw} on the Dirac equation. With this procedure, we can rigorously study the non-relativistic behavior
of spin-${1}/{2}$ particles, such as the electrons.
In this way, by looking at the ensuing Schr\"odinger equation
we can easily recognize the inertial and the gravitational mass.

In order to find the shape of $m_i$, one can think of an interaction between the electron and an external electric field, so that the Dirac equation which
includes the electromagnetic contribution is given by
\begin{equation}
\left(\slashed{p}-m_{0}-\frac{\alpha}{4\pi^{2}}\slashed{I}\right)\psi=e\Gamma_{\mu}A^{\mu}\psi.\label{eq:92}
\end{equation}
where $\alpha$ is the fine-structure constant, $A^{\mu}$ is the electromagnetic four-potential, with $A^{\mu}=\left(\phi,\mathbf{A}\right)$ ($\phi$ is the scalar potential, $\mathbf{A}$ the vector potential) and the quantity $I_\mu$ is  
\begin{equation}
I_{\mu}=2\int d^{3}k\frac{n_{B}\left(k\right)}{k_{0}}\frac{k_{\mu}}{\omega_{p}k_{0}-\mathbf{p}\cdot\mathbf{k}},\label{eq:89}
\end{equation}
with $k_{\mu}=\left(k_{0},\mathbf{k}\right)$. Clearly, $\omega_{p}$
and $\mathbf{p}$ participate in the dispersion relation
\begin{equation}
\omega_{p}=\sqrt{m_{0}^{2}+|\mathbf{p}|^{2}}.\label{eq:90}
\end{equation}
Furthermore, $n_{B}(k)$ represents the Bose-Einstein
distribution
\begin{equation}
n_{B}(k)=\frac{1}{e^{\beta k}-1},\label{eq:91}
\end{equation}
where $\beta={1}/{T}$.
Finally, $\Gamma_{\mu}$ turns out to be
\begin{equation}
\Gamma_{\mu}=\gamma_{\mu}\left(1-\frac{\alpha}{4\pi^{2}}\frac{I_{0}}{E}\right)+\frac{\alpha}{4\pi^{2}}I_{\mu}.\label{eq:94}
\end{equation}
Now, a Foldy\textendash Wouthuysen transformation allows us to deduce from 
Eq.~(\ref{eq:92}) its non-relativistic Schr\"odinger equation, that is
\begin{equation}
i\frac{\partial\psi_{s}}{\partial t}=\left[m_{0}+\frac{\alpha\pi T^{2}}{3m_{0}}+\frac{|\mathbf{p}|^{2}}{2\left(m_{0}+\frac{\alpha\pi T^{2}}{3m_{0}}\right)}+e\phi+\frac{\mathbf{p}\cdot\mathbf{A}+\mathbf{A}\cdot\mathbf{p}}{2\left(m_{0}+\frac{\alpha\pi T^{2}}{3m_{0}}\right)}+\ldots\right]\psi_{s},\label{eq:95}
\end{equation}
from which it is possible to detect the inertial mass
\begin{equation}
m_{i}=m_{0}+\frac{\alpha\pi T^{2}}{3m_{0}}.\label{eq:96}
\end{equation}
Interestingly, the difference between the
inertial mass of an electron at finite temperature and $m_{0}$ amounts to the thermal radiative correction of Eq.~(\ref{eq:96}). 

Similar considerations can be carried out also for the evaluation of the gravitational mass.
However, for this purpose we shall start from a different Dirac equation
that accounts for the presence of the gravitational interaction. To comply with the above reasoning, we can work in the weak-field limit, and note that the resulting Dirac equation for the electron yields
\begin{equation}
\left(\slashed{p}-m_{0}-\frac{\alpha}{4\pi^{2}}\slashed{I}\right)\psi=\frac{1}{2}h_{\mu\nu}\tau^{\mu\nu}\psi,\label{eq:97}
\end{equation}
where the fluctuations with respect to the flat background metric are denoted as $h_{\mu\nu}$
and with $\tau^{\mu\nu}$ being the renormalized stress-energy
tensor. According to Refs.~\cite{don}, we can take the metric tensor as written is isotropic coordinates, which entails $h_{\mu\nu}~=~2\,\phi_g\,\mathrm{diag}\left(1,1,1,1\right)$, where $\phi_g=-GM/r$ is the Newtonian potential.

As already done before, we now need to apply a Foldy\textendash Wouthuysen transformation to bring Eq.~(\ref{eq:97}) to another
Schr\"odinger equation
\begin{equation}
i\frac{\partial\psi_{s}}{\partial t}=\left[m_{0}+\frac{\alpha\pi T^{2}}{3m_{0}}+\frac{|\mathbf{p}|^{2}}{2\left(m_{0}+\frac{\alpha\pi T^{2}}{3m_{0}}\right)}+\left(m_{0}-\frac{\alpha\pi T^{2}}{3m_{0}}\right)\phi_{g}\right]\psi_{s},\label{eq:100}
\end{equation}
in which we can unambiguously identify the gravitational mass
\begin{equation}
m_{g}=\left(m_{0}-\frac{\alpha\pi T^{2}}{3m_{0}}\right).\label{eq:101}
\end{equation}
As expected, as long as $T=0$ there is absolutely no difference between $m_{g}$ and $m_{i}$, since they are
both represented by the renormalized mass.
Thus, we learn that only the existence of radiative corrections enables the violation of the EP.

As an additional step, from Eqs.~(\ref{eq:96}) and (\ref{eq:101}) we can compute the ratio $m_g/m_i$, which in the first-order approximation in $T^{2}$ turns out to be
\begin{equation}
\frac{m_{g}}{m_{i}}=1-\frac{2\alpha\pi T^{2}}{3m_{0}^{2}}.\label{eq:103}
\end{equation}
Physically speaking, EP violation is triggered by the spontaneous breaking of Lorentz symmetry due to the presence of a finite temperature, which renders the definition of an absolute motion
through the vacuum (i.e. the one at rest with the heat bath) a feasible scenario. 

Eq.~(\ref{eq:103}) is the starting point for the upcoming discussions, as it
will be examined in great detail and will be derived once again in the next Section, where we will rely on an alternative approach mainly centered around the gravitational sector rather than QFT.
As a matter of fact, it is licit to ask
whether $T$ can emerge from purely gravitational considerations without being concerned with the computation of radiative corrections.
Should this proposal be possible, we could readily perform analogous calculations
for several physical contexts that describe different spacetimes. 

\section{Equivalence Principle violation: a second approach}

As already argued, the quest for another route to reach the result~(\ref{eq:103})
by exclusively relying on GR is the main subject of the present Section. If we manage to reproduce the same ratio between $m_i$ and $m_g$ via a simpler approach, we could exploit such a formalism to investigate scenarios beyond GR, as for instance Extended Theories of Gravity. In what follows, we retrace the conceptual and mathematical steps contained in the seminal paper by Gasperini~\cite{gasp}.

As before, the objective is the analysis of a charged
test particle of renormalized mass at zero temperature $m_{0}$ in
thermal equilibrium with a photon heat bath in the low-temperature
limit $T\ll m_{0}$. To take into account the effects of $T$, the dispersion relation acquires an extra term which allows us to write~\cite{don}
\begin{equation}
E=\sqrt{m_{0}^{2}+|\mathbf{p}|^{2}+\frac{2}{3}\alpha\pi T^{2}}.\label{eq:104}
\end{equation}
At this point, let us introduce the stress-energy tensor $T^{\mu\nu}$
associated with the test particle, the world line of which can be enclosed in
a narrow ``world tube'' where $T^{\mu\nu}$ is non-vanishing. The conservation equations 
for the stress-energy tensor can be integrated over a three-dimensional
hypersurface $\Sigma$, thereby yielding
\begin{equation}
\int_{\Sigma}d^{3}x'\sqrt{-g}\,T^{\mu\nu}\left(x'\right)=\frac{p^{\mu}p^{\nu}}{E},\label{eq:105}
\end{equation}
where $p^{\mu}$ is the four-momentum and $E=p^{0}$ the energy
\begin{equation}
E=\int_{\Sigma}d^{3}x'\sqrt{-g}\,T^{00}\left(x'\right).\label{eq:106}
\end{equation}
The validity of these equations holds in the limit in which the world
tube's radius goes to zero~\cite{pap}.

A deeper analysis~\cite{gasp} demonstrates that Einstein's field equations shall be modified as well. To do that, we restrict our attention to the rest frame of the heat bath, where it is possible to observe that
\begin{equation}
\Xi^{\mu\nu}=T^{\mu\nu}-\frac{2}{3}\alpha\pi\frac{T^{2}}{E^{2}}\delta_{\;\;0}^{\mu}\delta_{\;\;0}^{\nu}T^{00},\label{eq:107}
\end{equation}
with $\Xi^{\mu\nu}$ being the counterpart of the Einstein tensor $G^{\mu\nu}$ that encompasses the thermal corrections due to the presence of a photon heat bath.
It is worth pointing out that Eq.~(\ref{eq:107}) is explicitly derived after the selection
of a preferred reference frame (i.e. the one at rest with the heat bath); this choice naturally leads to a Lorentz invariance violation of the finite temperature
vacuum. Indeed, by focusing on the flat tangent space, there is no notion of Minkowski vacuum anymore, since it has been replaced by
a thermal bath. For this reason, Lorentz group is no longer the symmetry
group of the local tangent space to the Riemannian manifold, even though we can reasonably require that general covariance still
holds there (for further details on general covariance, see Ref.~\cite{gencov}). Therefore, even though we are not precisely in the framework of GR, we can still safely assume that our reasoning is not affected by the aforementioned considerations~\cite{gasp}.

Now, since the examined problem deals with weak-field approximation
and quadratic thermal corrections in the low-temperature limit, the generalization
of Eq.~(\ref{eq:107}) to a curved spacetime can be thought of as
\begin{equation}
\Xi^{\mu\nu}=T^{\mu\nu}-\frac{2}{3}\alpha\pi\frac{T^{2}}{E^{2}}e_{\;\;\hat{0}}^{\mu}e_{\;\;\hat{0}}^{\nu}T^{\hat{0}\hat{0}},\label{eq:108}
\end{equation}
where $e_{\;\;\hat{0}}^{\mu}$ denotes the vierbein field and the hatted indexes are the ones related to the flat tangent space.

At this stage, before moving forward we must necessarily make a fundamental assumption: the implications of a non-vanishing temperature on the geometric structure of spacetime do not play a relevant r\^ole~\cite{gasp}. If the last statement is deemed as reliable, then we can promptly generalize Einstein's field equations as follows:
\begin{equation}
G^{\mu\nu}=\Xi^{\mu\nu}.\label{eq:109}
\end{equation}
Should we worked with the interplay between temperature and geometry, other contributions would have appeared from a relativistic study of the temperature, as $T$ would affect the surrounding spacetime; however, this kind of treatment lies beyond the scope of the present manuscript.

After that, we resort to the Bianchi identity
(i.e. $\nabla_{\nu}G^{\mu\nu}=0$) to write
\begin{equation}
\nabla_{\nu}T^{\mu\nu}=\nabla_{\nu}\left(\frac{2}{3}\alpha\pi\frac{T^{2}}{E^{2}}e_{\;\;\hat{0}}^{\mu}e_{\;\;\hat{0}}^{\nu}T^{\hat{0}\hat{0}}\right),\label{eq:110}
\end{equation}
which can be cast in the alternative form
\begin{equation}
\partial_{\nu}\left(\sqrt{-g}T^{\mu\nu}\right)+\Gamma_{\nu\alpha}^{\;\;\;\;\mu}\sqrt{-g}T^{\alpha\nu}=\partial_{\nu}\left(\sqrt{-g}\frac{2}{3}\alpha\pi\frac{T^{2}}{E^{2}}e_{\;\;\hat{0}}^{\mu}e_{\;\;\hat{0}}^{\nu}T^{\hat{0}\hat{0}}\right)+\frac{2}{3}\alpha\pi\Gamma_{\nu\alpha}^{\;\;\;\;\mu}\sqrt{-g}\frac{T^{2}}{E^{2}}e_{\;\;\hat{0}}^{\mu}e_{\;\;\hat{0}}^{\nu}T^{\hat{0}\hat{0}}.\label{eq:111}
\end{equation}
By denoting with $\overset{.}{x}^\mu\equiv dx^\mu/ds$, it can be shown~\cite{gasp} that Eq.~(\ref{eq:111}) can be further rephrased so as to give
\begin{equation}
\overset{..}{x}^{\mu}+\Gamma_{\alpha\nu}^{\;\;\;\;\mu}\overset{.}{x}^{\alpha}\overset{.}{x}^{\nu}=\frac{d}{ds}\left(\frac{2}{3}\alpha\pi\frac{T^{2}}{mE}e_{\;\;\hat{0}}^{\mu}\right)+\frac{2}{3}\alpha\pi\frac{T^{2}}{m^{2}}\Gamma_{\alpha\nu}^{\;\;\;\;\mu}e_{\;\;\hat{0}}^{\alpha}e_{\;\;\hat{0}}^{\nu},\label{eq:114}
\end{equation}
which can be manipulated by recalling that
\begin{equation}
E=m\overset{.}{x}^{\hat{0}}=m\overset{.}{x}^{\rho}e_{\rho}^{\;\;\hat{0}}.\label{eq:115}
\end{equation}
This substitution finally provides
\begin{equation}
\overset{..}{x}^{\mu}+\Gamma_{\alpha\nu}^{\;\;\;\;\mu}\overset{.}{x}^{\alpha}\overset{.}{x}^{\nu}=\frac{2}{3}\alpha\pi T^{2}\left[\frac{\overset{.}{x}^{\nu}\partial_{\nu}e_{\;\;\hat{0}}^{\mu}}{mE}-\frac{e_{\;\;\hat{0}}^{\mu}\left(\overset{..}{x}^{\nu}e_{\nu}^{\;\;\hat{0}}+\overset{.}{x}^{\nu}\overset{.}{x}^{\beta}\partial_{\beta}e_{\nu}^{\;\;\hat{0}}\right)}{E^{2}}+\frac{\Gamma_{\alpha\nu}^{\;\;\;\;\mu}e_{\;\;\hat{0}}^{\alpha}e_{\;\;\hat{0}}^{\nu}}{m^{2}}\right].\label{eq:116}
\end{equation}
Equation~(\ref{eq:116}) is nothing but a generalization of the geodesic equation that accounts for the presence of a non-vanishing temperature $T$. The significance of a similar extension can be easily conveyed by introducing a simple example; in fact, to begin with, we will consider the Schwarzschild solution. 

\subsection{The case of the Schwarzschild solution}

Let us now specialize Eq.~(\ref{eq:116}) in the context of the Schwarzschild metric, which in the spherical coordinates system can be written as
\begin{equation}
g_{\mu\nu}=\mathrm{diag}\left(e^\nu,-e^\lambda,-r^2,-r^2\mathrm{sin}^2\theta\right), \qquad e^{\nu}=e^{-\lambda}=1-2\phi_g=1-\frac{2GM}{r}.\label{eq:117}
\end{equation}
Furthermore, we recall that $\partial_{t}\phi_g=0$
and we additionally require the motion to take place along a radial trajectory (which means $\overset{.}{\vartheta}=\overset{.}{\varphi}=0$). Under these assumptions, we expect to essentially recover the same results obtained in Ref.~\cite{gasper}.

Bearing this in mind, we note that the non-vanishing vierbeins for the metric~(\ref{eq:117}) are represented by
\begin{equation}
e_{\;\;\hat{0}}^{0}=e^{-\frac{\nu}{2}};\;\;\;e_{\;\;\hat{1}}^{1}=e^{-\frac{\lambda}{2}}.\label{eq:119}
\end{equation}
To proceed, we need to evaluate the Christoffel symbols which explicitly enter in the geodesic equation for the temporal and the radial component. The relevant quantities that do not give zero identically are
\begin{equation}
\Gamma_{01}^{\;\;\;\;0}=\frac{\nu'}{2};\;\;\;\Gamma_{00}^{\;\;\;\;1}=\frac{\nu'}{2}\,e^{2\nu};\;\;\;\Gamma_{11}^{\;\;\;\;1}=-\frac{\nu'}{2},\label{eq:120}
\end{equation}
where $\nu=\mathrm{ln}\left(1-2\phi_g\right)$ and $\nu'={d\nu}/{dr}$.

Now, the geodesic equation for $\mu=0$ yields
\begin{equation}
\overset{..}{t}+\nu'\overset{.}{r}\overset{.}{t}=-\frac{2}{3}\alpha\pi T^{2}\left[\frac{\overset{.}{r}\nu'}{2mE}+\frac{\overset{..}{t}+\frac{\overset{.}{r}\overset{.}{t}\nu'}{2}}{E^{2}}e^{\frac{\nu}{2}}\right]e^{-\frac{\nu}{2}},\label{eq:122}
\end{equation}
but if one employs the relation $E=m\overset{.}{x}^{\hat{0}}=m\overset{.}{x}^{\alpha}e_{\alpha}^{\;\;\hat{0}}=m\,\overset{.}{t}\,e^{{\nu}/{2}}$,
Eq.~(\ref{eq:122}) can be cast as follows:
\begin{equation}
\overset{..}{t}+\nu'\overset{.}{r}\overset{.}{t}=-\frac{2\alpha\pi T^{2}}{3E^{2}}\left(\overset{..}{t}+\nu'\overset{.}{r}\overset{.}{t}\right),\label{eq:123}
\end{equation}
and since
$\overset{.}{\nu}=\nu'\overset{.}{r}$,
we then get
\begin{equation}
\left(1+\frac{2\alpha\pi T^{2}}{3E^{2}}\right)\left(\overset{..}{t}+\overset{.}{\nu}\overset{.}{t}\right)=0.\label{eq:125}
\end{equation}
The radial contribution can be computed by selecting $\mu=1$ in Eq.~(\ref{eq:116}), that is
\begin{equation}
\overset{..}{r}+\frac{\nu'}{2}\left(\overset{.}{t}^{2}e^{2\nu}-\overset{.}{r}^{2}\right)=\frac{2\alpha\pi T^{2}}{3m^{2}}\frac{e^{\nu}\nu'}{2},\label{eq:126}
\end{equation}
which can be manipulated so as to give
\begin{equation}
\overset{..}{r}+\frac{\nu'}{2}\left(\overset{.}{t}^{2}e^{\nu-\lambda}-\overset{.}{r}^{2}-\frac{2\alpha\pi T^{2}}{3m^{2}}e^{-\lambda}\right)=0.\label{eq:127}
\end{equation}
Equations (\ref{eq:125}) and (\ref{eq:127}) are exactly the same ones deduced in Ref.~\cite{gasper}, and they create a coupled system of
differential equations, which in general has non-trivial solutions. However, for the problem at hand
it is possible to come up with a handy relation between $\overset{.}{t}^{2}$
and $\overset{.}{r}^{2}$ that can be exploited to reach the final outcome.

As a matter of fact, Eq.~(\ref{eq:127}) can be rewritten as
\begin{equation}
2\overset{..}{r}-\overset{.}{r}^{2}\nu'+\overset{.}{t}^{2}\nu'e^{2\nu}-\frac{2\alpha\pi T^{2}}{3m^{2}}\nu'e^{\nu}=0,\label{eq:128}
\end{equation}
which can also be expressed as
\begin{equation}
e^{\nu}\frac{d}{dr}\left(e^{\lambda}\overset{.}{r}^{2}-e^{\nu}\overset{.}{t}^{2}-\frac{2\alpha\pi T^{2}}{3m^{2}}\nu\right)=0.\label{eq:135}
\end{equation}
The above equation clearly entails
\begin{equation}
e^{\lambda}\overset{.}{r}^{2}-e^{\nu}\overset{.}{t}^{2}-\frac{2\alpha\pi T^{2}}{3m^{2}}\nu=\mathrm{const}.\label{eq:136}
\end{equation}
The constant can be computed from the normalization condition
on four-velocity in the limit $\phi_g\rightarrow0$. Such a requirement
is a direct consequence of the independence of the geometric structure from the temperature; if this ansatz were relaxed, we would have a different result. For the moment, the
normalization of $\overset{.}{x}^{\mu}$ tells us that
\begin{equation}
\overset{.}{x}^{\mu}\overset{.}{x}_{\mu}=g_{\mu\nu}\overset{.}{x}^{\mu}\overset{.}{x}^{\nu}=1,\label{eq:137}
\end{equation}
or explicitly
\begin{equation}
e^{\lambda}\overset{.}{r}^{2}-e^{\nu}\overset{.}{t}^{2}=-1,\label{eq:138}
\end{equation}
as we are assuming radial motion.

In the limit of vanishing gravitational field (i.e. $\nu,\lambda\rightarrow 0$ as $r\rightarrow\infty$),
Eq.~(\ref{eq:138}) becomes
\begin{equation}
\overset{.}{r}_{\infty}^{2}-\overset{.}{t}_{\infty}^{2}=-1.\label{eq:139}
\end{equation}
Such an expression can be adapted to Eq.~(\ref{eq:136}); thus, we are left with
\begin{equation}
e^{\lambda}\overset{.}{r}^{2}-e^{\nu}\overset{.}{t}^{2}-\frac{2\alpha\pi T^{2}}{3m^{2}}\nu=-1.\label{eq:140}
\end{equation}
At this point, we use the assumption of the weak-field limit. Within this domain and by virtue of Eq.~(\ref{eq:140}), it is straightforward to observe that Eq.~(\ref{eq:127}) can be cast as
\begin{equation}
\overset{..}{r}=-\frac{GM}{r^{2}}\left(1-\frac{2\alpha\pi T^{2}}{3m^{2}}\right),\label{eq:145}
\end{equation}
and if one considers the first-order approximation in $T^{2}$ just like in
QFT considerations, we obtain
\begin{equation}\nonumber
\frac{m_{g}}{m_{i}}=1-\frac{2\alpha\pi T^{2}}{3m_{0}^{2}},\label{eq:146}
\end{equation}
which is the same outcome contained in Eq.~(\ref{eq:103}).

With the above achievement, we have established a one-to-one correspondence between the QFT approach and the one formulated in Ref.~\cite{gasp}, even though the starting assumptions and the ensuing developments are totally different. However, both of the two studies rely on finite-temperature considerations, which is the fundamental feature that allows the ratio ${m_{g}}/{m_{i}}$ to deviate from unity.

On the other hand, the modified geodesic equation
is a general result, and as such it can be investigated in any physical scenario;
the only quantity required for calculations is the metric tensor. By virtue of its knowledge,
Eq.~(\ref{eq:116}) can be specialized to deduce the differential equations
from which one can compute the ratio between the inertial and the gravitational mass, with the aim of
revealing any potential trace of EP violation.

\subsection{Generic diagonal metrics}

In the following, we derive a general result which holds true for any given diagonal metric tensor cast in spherical coordinates and that depends only on $r$. According to this prescription, we can write $g_{\mu\nu}$ as
\begin{equation}
g_{\mu\nu}=\mathrm{diag}\left(A(r),-\frac{1}{B(r)},-r^2,-r^2\mathrm{sin}^2\theta\right).
\end{equation}
To move forward, we need to introduce the non-vanishing Christoffel symbols and vierbein fields, which in this case are represented by
\begin{equation}
e_{\;\;\hat{0}}^{0}=\left(A\right)^{-\frac{1}{2}};\;\;\;e_{\;\;\hat{1}}^{1}=\left(B\right)^{\frac{1}{2}},\label{eq:263}
\end{equation}
\begin{equation}
\Gamma_{01}^{\;\;\;\;0}=\frac{\partial_{r}A(r)}{2A(r)};\;\;\;\Gamma_{00}^{\;\;\;\;1}=\frac{B(r)}{2}\partial_{r}A(r);\;\;\;\Gamma_{11}^{\;\;\;\;1}=\frac{B(r)}{2}\partial_{r}\left(\frac{1}{B(r)}\right).\label{eq:a22}
\end{equation}
As before, we have tacitly required radial motion, for which $\overset{.}{\theta}=\overset{.}{\varphi}=0$.

The development of the computation exactly reflects the same steps exhibited in the previous Subsection. Indeed, we first have to analyze the geodesic equation for the temporal coordinate; this procedure gives
\begin{equation}
\overset{..}{t}+\overset{.}{r}\overset{.}{t}\frac{\partial_{r}A}{A}=\frac{2}{3}\alpha\pi T^{2}\left[-\frac{\overset{.}{r}}{2mE}\frac{\partial_{r}A}{\left(A\right)^{\frac{3}{2}}}-\frac{1}{E^{2}}\left(\overset{..}{t}+\overset{.}{r}\overset{.}{t}\frac{\partial_{r}A}{2A}\right)\right],\label{eq:264}
\end{equation}
but since $E=m\,\overset{.}{t}\,e_{0}^{\;\;\hat{0}}=m\,\overset{.}{t}\,\sqrt{A}$,
Eq.~(\ref{eq:264}) is also equal to 
\begin{equation}
\overset{..}{t}+\overset{.}{r}\overset{.}{t}\frac{\partial_{r}A}{A}=-\frac{2\alpha\pi T^{2}}{3E^{2}}\left[\overset{..}{t}+\overset{.}{r}\overset{.}{t}\frac{\partial_{r}A}{A}\right],\label{eq:265}
\end{equation}
or 
\begin{equation}
\left(1+\frac{2\alpha\pi T^{2}}{3E^{2}}\right)\left(\overset{..}{t}+\overset{.}{r}\overset{.}{t}\frac{\partial_{r}A}{A}\right)=0,\label{eq:266}
\end{equation}
which recovers Eq.~(\ref{eq:125}) with the identification $A=e^{\nu}$.

On the other hand, when dealing with the radial geodesic equation we observe that
\begin{equation}
\overset{..}{r}+\overset{.}{r}^{2}\frac{B}{2}\partial_{r}\left(\frac{1}{B}\right)+\overset{.}{t}^{2}\frac{B\partial_{r}A}{2}=\frac{2\alpha\pi T^{2}}{3m^{2}}\frac{B}{2}\frac{\partial_{r}A}{A},\label{eq:267}
\end{equation}
or
\begin{equation}
\overset{..}{r}+\frac{B}{2}\left[\overset{.}{t}^{2}\partial_{r}A+\overset{.}{r}^{2}\partial_{r}\left(\frac{1}{B}\right)-\frac{2\alpha\pi T^{2}}{3m^{2}}\frac{\partial_{r}A}{A}\right]=0,\label{eq:268}
\end{equation}
which again returns Eq.~(\ref{eq:127}) with the appropriate choice for the functions 
$A$ and $B$.

At this point, from the first relation we can straightforwardly obtain
\begin{equation}
\frac{d}{ds}\left(\mathrm{ln}\,\overset{.}{t}\right)=-\frac{d}{ds}\left(\mathrm{ln}\,A\right),\label{eq:269}
\end{equation}
which tells us that
\begin{equation}
\overset{.}{t}=\frac{1}{A}.\label{eq:270}
\end{equation}
Concerning the radial equation, we can cast it as a total derivative with respect to $r$, namely
\begin{equation}
\frac{B}{2}\frac{d}{dr}\left[\frac{\overset{.}{r}^{2}}{B}-\overset{.}{t}^{2}A-\frac{2\alpha\pi T^{2}}{3m^{2}}\mathrm{ln}\left(A\right)\right]=0.\label{eq:276}
\end{equation}
Assuming that the normalization condition introduced above is still valid, we can deduce
\begin{equation}
\frac{\overset{.}{r}^{2}}{B}-\overset{.}{t}^{2}A-\frac{2\alpha\pi T^{2}}{3m^{2}}\mathrm{ln}\left(A\right)=-1,\label{eq:277}
\end{equation}
which is the generalization of Eq.~(\ref{eq:140}). After some algebra, Eq.~(\ref{eq:277}) becomes
\begin{equation}
\overset{.}{r}^{2}=B\left(\frac{1}{A}+\frac{2\alpha\pi T^{2}}{3m^{2}}\mathrm{ln}\left(A\right)-1\right).\label{eq:278}
\end{equation}
Finally, by resorting to Eqs.~(\ref{eq:270}) and (\ref{eq:278}),
we see that Eq.~(\ref{eq:268}) can be written as
\begin{equation}
\overset{..}{r}=-\frac{B}{2}\left[\frac{\partial_{r}A}{A^{2}}-\left(\frac{1}{A}-1\right)\frac{\partial_{r}B}{B}-\frac{2\alpha\pi T^{2}}{3m^{2}}\left(\frac{\partial_{r}A}{A}+\frac{\mathrm{ln}\left(A\right)\partial_{r}B}{B}\right)\right].\label{eq:279}
\end{equation}
Equation~(\ref{eq:279}) represents the starting point for the next Section, where we employ known solutions for the metric tensor in spherical coordinates associated with Extended Theories of Gravity to properly quantify the amount of EP violation. 

\section{Equivalence Principle violation in curvature based Extended Gravity}

As already anticipated, in what follows we stick to Eq.~(\ref{eq:279}) to explore the interplay between the implications of several Extended Theories of Gravity and the EP. For all the upcoming examples, we will see how the departure from Einstein's GR unavoidably results in EP violation. 

\subsection{The Brans-Dicke Theory}

The Brans-Dicke theory~\cite{bd} is the most famous scalar-tensor theory of gravity. As their name suggests, these theories predict the existence of an auxiliary scalar field which, together with the usual metric tensor, mediates the gravitational interaction. For the case under examination, such an auxiliary field is not interpreted as a new particle degree of freedom, but is rather related to the idea of a shifting Newton constant $G$. For this reason, several restriction must be required so as to let the model be compatible with the outcomes of gravitational experiments.
 
Now, let us define the Brans-Dicke action~\cite{bd} as
\begin{equation}
S_{BD}=\int d^4x\sqrt{-g}\left(\varphi R-\omega\frac{1}{\varphi}g^{\mu\nu}\partial_{\mu}\varphi\partial_{\nu}\varphi+\mathfrak{L}_\mathrm{matter}\left(\psi\right)\right).\label{eq:50}
\end{equation}
where $\omega$ is the free parameter of the theory and
\begin{equation}
\varphi=\frac{1}{16\pi G_{eff}},\label{eq:51}
\end{equation}
is a scalar field indicating an ``effective''
gravitational constant.
However, $\varphi$ must be spatially uniform and it must slowly change
with the cosmic time, otherwise the Brans-Dicke theory clashes with experimental data.

For this model, it is possible
to find a Schwarzschild-like solution, which in isotropic coordinates reads~\cite{bd}
\begin{equation}
ds^{2}=e^{v}dt^{2}-e^{u}\left[dr^{2}+r^{2}\left(d\vartheta^{2}+\sin^{2}\vartheta\,d\varphi^{2}\right)\right],\label{eq:eff57}
\end{equation}
where
\begin{equation}
e^{v}=e^{2\alpha_{0}}\left(\frac{1-\frac{B}{r}}{1+\frac{B}{r}}\right)^{\frac{2}{\lambda}}, \qquad e^{u}=e^{2\beta_{0}}\left(1+\frac{B}{r}\right)^{4}\left(\frac{1-\frac{B}{r}}{1+\frac{B}{r}}\right)^{\frac{2\left(\lambda-C-1\right)}{\lambda}},\label{eq:eff58}
\end{equation}
with $\alpha_{0}$, $\beta_{0}$, $B$, $C$ and $\lambda$ being functions of $\omega$.
In this framework, the solution for the scalar field is given by
\begin{equation}
\phi=\phi_{0}\left(\frac{1-\frac{B}{r}}{1+\frac{B}{r}}\right)^{-\frac{C}{\lambda}},\label{eq:eff61}
\end{equation}
with $\phi_{0}=\mathrm{constant}$. 

Although the above solution is not written in spherical coordinates\footnote{We remark that the analysis of the Brans-Dicke model has been carried out for the sake of continuity with the seminal paper on the topic~\cite{original}.}, one can still evaluate the geodesic equations and come up with a suitable expression where to identify the quantity $m_g/m_i$. According to Ref.~\cite{original}, if the procedure explained in the previous Section were applied for the current model, it would yield
\begin{equation}
\overset{..}{r}=-\frac{v'}{2}\left\{ 1+\left(e^{-v}-1\right)\left(\frac{\lambda B}{r}-C\right)-\frac{2\alpha\pi T^{2}}{3m^{2}}\left[1+v-\left(\frac{\lambda B}{r}-C\right)v\right]\right\} e^{-u}.\label{eq:eff279}
\end{equation}
It is worth pointing out that the radiative correction is not the only contribution to the ratio ${m_{g}}/{m_{i}}$,
as there is another term which only depends on $\omega$ and that
correctly recovers GR as long as $\omega\rightarrow\infty$.
By further investigating the latter factor, in principle one can impose a lower bound to the free parameter
of the Brans-Dicke theory. Indeed, by requiring $|({m_{g}-m_{i}})/{m_{i}}|<10^{-14}$~\cite{bae} and working in the weak-field limit (but when thermal corrections are negligible), one can easily constrain $\omega$ . 
To achieve the desired outcome, we have to regard the quantities appearing in the metric tensor~(\ref{eq:eff57}) as functions of $\omega$ for weak gravitational fields. Such a study has been performed in Ref.~\cite{bar}, according to which we have
\begin{equation}
\alpha_{0}=\beta_{0}=0;\;\;C=-\frac{1}{2+\omega};\;\;B=\frac{GM\lambda}{2};\;\;\lambda=\sqrt{\frac{2\omega+3}{2\omega+4}}.\label{eq:299}
\end{equation}
Now, we can immediately write down $B$ in terms of the Newtonian potential $\phi_{g}$, that is
\begin{equation}
B=-\frac{\lambda r\phi_{g}}{2},\label{eq:300}
\end{equation}
and expand the function $e^{-v}$ as
\begin{equation}
e^{-v}=\left(\frac{1-\frac{\lambda\phi_{g}}{2}}{1+\frac{\lambda\phi_{g}}{2}}\right)^{-\frac{2}{\lambda}}\sim1+2\phi_{g}.\label{eq:301}
\end{equation}
Since we consider only the linearized case, we shall stop at $\mathcal{O}(\phi_g)$, for which we note that
\begin{equation}
\lf|\frac{m_g-m_i}{m_i}\ri|=\frac{2\phi_{g}}{2+\omega},\label{eq:302}
\end{equation}
and thus from ${2\phi_{g}}/({2+\omega})<10^{-14}$ we obtain
\begin{equation}
\omega>\frac{2GM}{r}\cdot10^{14}.\label{eq:305}
\end{equation}
For the gravitational field of the Earth, we have that
$M_{\oplus}=5.97\cdot10^{24}$ Kg and $R_{\oplus}=6.37\cdot10^{6}$ m, and in turn
\begin{equation}
\omega>1.40\cdot10^{5},\label{eq:308}
\end{equation}
that is close to an experimental bound recently achieved~\cite{wy} which is equal to $\omega>3\cdot10^{5}$.

\subsection{The Standard Model Extension}

The Standard Model Extension (SME)~\cite{sme1} is a string-theoretical effective field theory that broadens the domain of validity of the Standard Model (SM) by accounting for the spontaneous breaking of Lorentz symmetry. Such a scenario is achieved by introduction Lorentz-violating operators that are contracted with the usual SM fields, thus in principle allowing for new physics phenomenology even at the currently reachable scales.

As we are concerned with the gravity sector only~\cite{sme2}, we can write down the most general Lorentz-violating action as 
\be\label{lvac}
S_{LV}=\frac{1}{16\pi G}\int d^4x\sqrt{-g}\lf[(1-u)R+s^{\mu\nu}R^T_{\mu\nu}+t^{\mu\nu\rho\lambda}C_{\mu\nu\rho\lambda}\ri]\,,
\ee
with $R$ being the Ricci scalar, $R^T_{\mu\nu}$ the trace-free Ricci tensor, $C_{\mu\nu\rho\lambda}$ the Weyl conformal tensor whereas $u$, $s^{\mu\nu}$ and $t^{\mu\nu\rho\lambda}$ are the Lorentz-violating fields. 
In the non-relativistic limit, it is possible to show~\cite{sme3} that the dominant SME correction to the Schwarzschild solution is given by
\be\label{linel}
ds^2=F(r)dt^2-\frac{1}{F(r)}dr^2-r^2d\Omega^2\,, \qquad F(r)=1-\frac{2GM}{r}\lf[1+\bar{s}^{ij}\chi_{ij}(\theta,\varphi)\ri]\,,
\ee
where $\bar{s}^{ij}$ are the vacuum expectation values of $s^{ij}$ and $\chi_{ij}(\theta,\varphi)$ functions of the angular coordinates. Note that the inequality $\chi_{ij}(\theta,\varphi)\leq1$ is valid for any value of $\theta$ and $\varphi$~\cite{sme3,ls}. The comparison with Eq.~(\ref{eq:279}) immediately implies that $A(r)=B(r)=F(r)$, and therefore we can resort to the outcome obtained in the previous Section to observe that 
\be\label{res2}
\overset{..}{r}=-\frac{GM}{r^2}\lf[1+\bar{s}^{ij}\chi_{ij}(\theta,\varphi)-\frac{2\al\pi T^2}{3m^2}\lf(1+\bar{s}^{ij}\chi_{ij}(\theta,\varphi)\ri)\lf(1+\ln\lf[F(r)\ri]\ri)\ri]\,.
\ee
If we assume the radiative corrections to be negligible (as done for the Brans-Dicke framework), we see that the breakdown of Lorentz symmetry results in a violation of the EP even at $T=0$, as the difference between the inertial and the gravitational mass is related to the vacuum expectation value of Lorentz-violating fields. Furthermore, we can put a bound on the Lorentz-violating parameters by requiring consistency with the experimental data. As a matter of fact, from Eq.~(\ref{res2}) we see that 
\be\label{ratio2}
\frac{m_g}{m_i}=1+\bar{s}^{ij}\chi_{ij}(\theta,\varphi)\,,
\ee
from which we can derive an upper limit for $\bar{s}^{ij}$ by means of Ref.~\cite{bae}. Indeed, by assuming $|\chi_{ij}|\simeq1$~\cite{ls}, we deduce that 
\be\label{bou2}
\lf|\frac{m_g-m_i}{m_i}\ri|=|\bar{s}^{ij}|<10^{-14}\,,
\ee
which is of the same order of the experimental constraints available for this quantity~\cite{smerev}.

\subsection{Conformal gravity}

Conformal gravity is an extension of Einstein's GR that leaves the gravitational action invariant under conformal transformations. From a quantum field theoretical perspective, such a model predicts the appearance of higher-derivative terms, which provide a partial cure to the bad UV divergences of general relativity. From a cosmological viewpoint, it has been recently shown~\cite{mann} that conformal gravity is capable of describing the rotation curves of many dwarf galaxies without invoking dark matter. 

The starting point is given by the action
\be\label{cact}
S_C=-\al\int d^4x\sqrt{-g}\,C_{\mu\nu\rho\lambda}C^{\mu\nu\rho\lambda}\,,
\ee
with $\al$ being a dimensionless quantity. In the context of conformal gravity, one can compute an exact Schwarzschild-like solution, as done for the first time in Ref.~\cite{mann2}; the ensuing line element reads
\be\label{conmet}
ds^2=H(r)dt^2-\frac{1}{H(r)}dr^2-r^2d\Omega^2\,, \qquad H(r)=1-\frac{\beta(2-3\beta\gamma)}{r}-3\beta\gamma+\gamma r-kr^2\,,
\ee
where $\beta$, $\gamma$ and $k$ are parameters that must be fixed by experiments. Note that, in order to recover GR in a suitable limit, we must require $\beta(2-3\beta\gamma)/2=GM$. Furthermore, an interesting observation is in order here: a fine tuning of these factors also allows to recover several solutions belonging to the class of $f(R)$ models (see for instance the ones stemming from the considerations carried out in Refs.~\cite{capo}).

At this point, the identification $A(r)=B(r)=H(r)$ lets us easily compute Eq.~(\ref{eq:279}) for conformal gravity, which turns out to be
\be\label{res3}
\overset{..}{r}=-\frac{GM}{r^2}\lf[1+\frac{\gamma r^2}{2GM}-\frac{kr^3}{GM}-\frac{2\al\pi T^2}{3m^2}\lf(1+\frac{\gamma r^2}{2GM}-\frac{kr^3}{GM}\ri)\lf(1+\ln\lf[H(r)\ri]\ri)\ri]\,.
\ee
Again, by neglecting radiative corrections we observe that conformal gravity is at odds with the EP, as the ratio of the gravitational and the inertial mass is explicitly given by
\be\label{ratio3}
\frac{m_g}{m_i}=1+\frac{\gamma r^2}{2GM}-\frac{kr^3}{GM}\,.
\ee
By resorting to $M_\oplus$ and $R_\oplus$, we can also put a constraint on the combination of $\gamma$ and $k$, which yields
\be\label{bou3}
\lf|\frac{\gamma}{2}-kR_\oplus\ri|\lesssim10^{-13}\,.
\ee
However, since these quantities have been at the basis of a significant number of investigations revolving around the rotation curves of galaxies~\cite{mann}, the above bound is not the most stringent one that can be found in literature. Nevertheless, it is worth remarking that, because of Eq.~(\ref{ratio3}), EP violation becomes increasingly relevant as we move towards cosmological scales, thereby being potentially involved in the mechanisms that are able to explain the experimental evidences without relying on dark matter.

\section{Concluding remarks}
In this paper, we have studied the violation of the EP arising from the existence of a non-zero temperature. A similar result sinks its roots in thermal field theory, where radiative corrections are responsible for the difference between the inertial and the gravitational mass. Despite this, the very same achievement can be obtained without QFT considerations; indeed, we have seen how the influence of the temperature on gravitating quantum systems amounts to modify the geodesic equation. By virtue of the latter simplified procedure, in principle one can analyze different physical settings, as for example Extended Theories of Gravity and particle propagation in curved backgrounds\footnote{In particular, we remark that the investigation of neutrino physics in the presence of gravitational fields is intimately tied to the EP violation, as the vast literature on this topic suggests (see Refs.~\cite{lc} and therein). For instance, note that, even in the presence of standard gravity described by GR, the ratio $m_i/m_g$ is not unity, but it depends on the mixing properties~\cite{nonrel}.}. As a matter of fact, by working in the former context we have seen how several extended models of gravity (i.e. Brans-Dicke, Standard Model Extension and conformal gravity) exhibit a violation of the EP also in the absence of radiative corrections. Such an occurrence can be quantitatively substantiated by transferring the available bounds for the ratio $|(m_g-m_i)/m_i|$ onto the free parameters of the theory. To summarize the experimental developments obtained so far, in Fig.~1 we show the most relevant tests associated with WEP and the ensuing bound on the parameter of interest. It is worth stressing that the latest constraint achieved with the MICROSCOPE initiative still needs to be confirmed with greater precision~\cite{micro}. For more details on the tests, the reader can consult Refs.~\cite{will,weprev}.
Along this line, it is evident that, by refining the experimental sensitivity, we can have access to more stringent constraints for the extended models of gravity; fortunately, such technological improvements may be available in the next few years~\cite{na}. On a final note, it is appropriate to point out that the same protocol described above has already been employed in conjunction with the Casimir effect to probe the implications of theories beyond GR~\cite{lamb}. 
Other tests could come from ground based gyroscopes adopting ring lasers like the experiment GINGER \cite{Ginger1,Ginger2}.  These experimental researches are very active in this moment. Finally, we want to point out that the above protocol can be adopted not only for curvature based extensions of GR but also for affine theories based on torsion invariants \cite{cai,Salucci} and non-metric theories \cite{Trinity}. In other words, a possible detection of EP violation could be a fundamental tool to discriminate among concurring theories of gravity.

\begin{figure}[ht]
\centering
\includegraphics[width=1\textwidth]{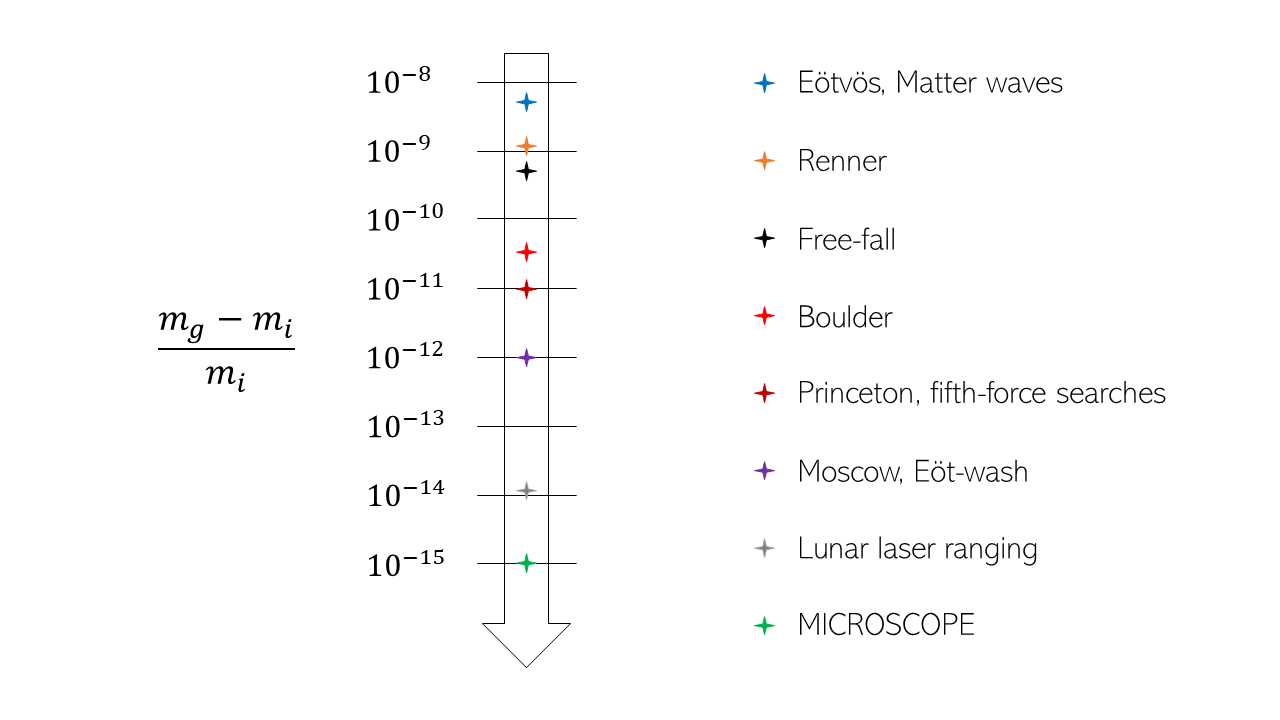}
\caption{Different tests of the Equivalence Principle and the ensuing values for the ratio $(m_g-m_i)/m_i$. This fraction can be straightforwardly deduced from the E\"otv\"os number, which is the physical quantity detected by experiments.}
\end{figure}

\end{document}